\documentclass[conference, 10pt]{IEEEtran}

\ifCLASSOPTIONcompsoc
 \usepackage[caption=false,font=normalsize,labelfont=sf,textfont=sf]{subfig}
\else
 \usepackage[caption=false,font=footnotesize]{subfig}
\fi

\usepackage{graphicx}
\usepackage{color}
\usepackage{placeins}
\usepackage{float}
\usepackage{tabularx,colortbl}
\usepackage[nolist]{acronym}
\usepackage[T1]{fontenc}

\begin{document}
\sloppy
\begin{acronym}

\acro{2D}{Two Dimensions}%
\acro{2G}{Second Generation}%
\acro{3D}{Three Dimensions}%
\acro{3G}{Third Generation}%
\acro{3GPP}{Third Generation Partnership Project}%
\acro{3GPP2}{Third Generation Partnership Project 2}%
\acro{4G}{Fourth Generation}%
\acro{5G}{Fifth Generation}%

\acro{AI}{Artificial Intelligence}%
\acro{AoA}{Angle of Arrival}%
\acro{AoD}{Angle of Departure}%
\acro{AR}{Augmented Reality}%
\acro{AP}{Access Point}
\acro{AE}{Antenna Element}
\acro{AC}{Anechoic Chamber}
\acro{AUT}{Antenna Under Test}

\acro{BER}{Bit Error Rate}%
\acro{BPSK}{Binary Phase-Shift Keying}%
\acro{BRDF}{ Bidirectional Reflectance Distribution Function}%
\acro{BS}{Base Station}%

\acro{CA}{Carrier Aggregation}%
\acro{CDF}{Cumulative Distribution Function}%
\acro{CDM}{Code Division Multiplexing}%
\acro{CDMA}{Code Division Multiple Access}%
\acro{CPU} {Central Processing Unit}
\acro{CUDA}{Compute Unified Device Architecture}
\acro{CDF}{Cumulative Distribution Function}
\acro{CI}{Confidence Interval}
\acro{CVRP}{Constrained-View Radiated Power}
\acro{CATR}{Compact Antenna Test Range}
 
\acro{D2D}{Device-to-Device}%
\acro{DL}{Down Link}%
\acro{DS}{Delay Spread}%
\acro{DAS}{Distributed Antenna System}
\acro{DKED}{double knife-edge diffraction}
\acro{DUT}{Device Under Test}


\acro{EDGE}{Enhanced Data rates for GSM Evolution}%
\acro{EIRP}{Equivalent Isotropic Radiated Power}%
\acro{eMBB}{Enhanced Mobile Broadband}%
\acro{eNodeB}{evolved Node B}%
\acro{ETSI}{European Telecommunications Standards Institute}%
\acro{ER}{Effective Roughness}%
\acro{E-UTRA}{Evolved UMTS Terrestrial Radio Access}%
\acro{E-UTRAN}{Evolved UMTS Terrestrial Radio Access Network}%
\acro{EF}{Electric Field}

\acro{FDD}{Frequency Division Duplexing}%
\acro{FDM}{Frequency Division Multiplexing}%
\acro{FDMA}{Frequency Division Multiple Access}%
\acro{FoM}{Figure of Merit}
\acro{FoV}{Field of View}
\acro{GI}{Global Illumination} %
\acro{GIS}{Geographic Information System}%
\acro{GO}{Geometrical Optics} %
\acro{GPU}{Graphics Processing Unit}%
\acro{GPGPU}{General Purpose Graphics Processing Unit}%
\acro{GPRS}{General Packet Radio Service}%
\acro{GSM}{Global System for Mobile Communication}%
\acro{GNSS}{Global Navigation Satellite System}%
\acro{H2D}{Human-to-Device}%
\acro{H2H}{Human-to-Human}%
\acro{HDRP}{High Definition Render Pipeline}
\acro{HSDPA}{High Speed Downlink Packet Access}
\acro{HSPA}{High Speed Packet Access}%
\acro{HSPA+}{High Speed Packet Access Evolution}%
\acro{HSUPA}{High Speed Uplink Packet Access}
\acro{HPBW}{Half-Power Beamwidth}

\acro{IEEE}{Institute of Electrical and Electronic Engineers}%
\acro{InH}{Indoor Hotspot} %
\acro{IMT} {International Mobile Telecommunications}%
\acro{IMT-2000}{\ac{IMT} 2000}%
\acro{IMT-2020}{\ac{IMT} 2020}%
\acro{IMT-Advanced}{\ac{IMT} Advanced}%
\acro{IoT}{Internet of Things}%
\acro{IP}{Internet Protocol}%
\acro{ITU}{International Telecommunications Union}%
\acro{ITU-R}{\ac{ITU} Radiocommunications Sector}%
\acro{IS-95}{Interim Standard 95}%
\acro{IES}{Inter-Element Spacing}


\acro{KPI}{Key Performance Indicator}%
\acro{K-S}{Kolmogorov-Smirnov}

\acro{LB} {Light Bounce}
\acro{LIM}{Light Intensity Model}%
\acro{LoS}{line of sight}%
\acro{LTE}{Long Term Evolution}%
\acro{LTE-Advanced}{\ac{LTE} Advanced}%
\acro{LSCP}{Lean System Control Plane}%
\acro{LSI} {Light Source Intensity}

\acro{M2M}{Machine-to-Machine}%
\acro{MatSIM}{Multi Agent Transport Simulation}
\acro{METIS}{Mobile and wireless communications Enablers for Twenty-twenty Information Society}%
\acro{METIS-II}{Mobile and wireless communications Enablers for Twenty-twenty Information Society II}%
\acro{MIMO}{Mul\-ti\-ple-In\-put Mul\-ti\-ple-Out\-put}
\acro{mMIMO}{massive MIMO}%
\acro{mMTC}{massive Machine Type Communications}%
\acro{mmW}{millimeter-wave}%
\acro{MU-MIMO}{Multi-User MIMO}
\acro{MMF}{Max-Min Fairness}
\acro{MKED}{Multiple Knife-Edge Diffraction}
\acro{MF}{Matched Filter}
\acro{mmWave}{Millimeter Wave}

\acro{NFV}{Network Functions Virtualization}%
\acro{NLoS}{non line of sight}%
\acro{NR}{New Radio}%
\acro{NRT}{Non Real Time}%
\acro{NYU}{New York University}%
\acro{N75PRP}{Near-75-degrees Partial Radiated Power}%
\acro{NHPRP}{Near-Horizon Partial Radiated Power}%

\acro{O2I}{Outdoor to Indoor}%
\acro{O2O}{Outdoor to Outdoor}%
\acro{OFDM}{Orthogonal Frequency Division Multiplexing}%
\acro{OFDMA}{Or\-tho\-go\-nal Fre\-quen\-cy Di\-vi\-sion Mul\-ti\-ple Access}
\acro{OtoI}{Outdoor to Indoor}%
\acro{OTA}{Over-The-Air}

\acro{PDF}{Probability Distribution Function}
\acro{PDP}{Power Delay Profile}
\acro{PHY}{Physical}%
\acro{PLE}{Path Loss Exponent}
\acro{PRP}{Partial Radiated Power}

\acro{QAM}{Quadrature Amplitude Modulation}%
\acro{QoS}{Quality of Service}%

\acro{RCSP}{Receive Signal Code Power}
\acro{RAN}{Radio Access Network}%
\acro{RAT}{Radio Access Technology}%

\acro{RAN}{Radio Access Network}%
\acro{RMa}{Rural Macro-cell}%
\acro{RMSE} {Root Mean Square Error}
\acro{RSCP}{Receive Signal Code Power}%
\acro{RT}{Ray Tracing}
\acro{RX}{receiver}
\acro{RMS}{Root Mean Square}
\acro{Random-LOS}{Random Line-Of-Sight}
\acro{RF}{Radio Frequency}
\acro{RC}{Reverberation Chamber}
\acro{RIMP}{Rich Isotropic Multipath}

\acro{SB} {Shadow Bias}
\acro{SC}{small cell}
\acro{SDN}{Software-Defined Networking}%
\acro{SGE}{Serious Game Engineering}%
\acro{SF}{Shadow Fading}%
\acro{SIMO}{Single Input Multiple Output}%
\acro{SINR}{Signal to Interference plus Noise Ratio}
\acro{SISO}{Single Input Single Output}%
\acro{SMa}{Suburban Macro-cell}%
\acro{SNR}{Signal to Noise Ratio}
\acro{SU}{Single User}%
\acro{SUMO}{Simulation of Urban Mobility}
\acro{SS} {Shadow Strength}


\acro{TDD}{Time Division Duplexing}%
\acro{TDM}{Time Division Multiplexing}%
\acro{TD-CDMA}{Time Division Code Division Multiple Access}%
\acro{TDMA}{Time Division Multiple Access}%
\acro{TX}{transmitter}
\acro{TZ}{Test Zone}
\acro{TRP}{Total Radiated Power}


\acro{UAV}{Unmanned Aerial Vehicle}%
\acro{UE}{User Equipment}%
\acro{UI}{User Interface}
\acro{UHD}{Ultra High Definition}
\acro{UL}{Uplink}%
\acro{UMa}{Urban Macro-cell}%
\acro{UMi}{Urban Micro-cell}%
\acro{uMTC}{ultra-reliable Machine Type Communications}%
\acro{UMTS}{Universal Mobile Telecommunications System}%
\acro{UPM}{Unity Package Manager}
\acro{UTD}{Uniform Theory of Diffraction} %
\acro{UTRA}{{UMTS} Terrestrial Radio Access}%
\acro{UTRAN}{{UMTS} Terrestrial Radio Access Network}%
\acro{URLLC}{Ultra-Reliable and Low Latency Communications}%
\acro{UHRP}{Upper Hemisphere Radiated Power}%

\acro{V2V}{Vehicle-to-Vehicle}%
\acro{V2X}{Vehicle-to-Everything}%
\acro{VP}{Visualization Platform}%
\acro{VR}{Virtual Reality}%
\acro{VNA}{Vector Network Analyzer}
\acro{VIL}{Vehicle-in-the-loop}

\acro{WCDMA}{Wideband Code Division Multiple Access}%
\acro{WINNER}{Wireless World Initiative New Radio}%
\acro{WINNER+}{Wireless World Initiative New Radio +}%
\acro{WiMAX}{Worldwide Interoperability for Microwave Access}%
\acro{WRC}{World Radiocommunication Conference}%

\acro{xMBB}{extreme Mobile Broadband}%

\acro{ZF}{Zero Forcing}

\end{acronym}

%
\title{Constrained FoV Radiated Power as a Figure of Merit of Phased Arrays}

%

\author{\IEEEauthorblockN{Alejandro Antón Ruiz}
\IEEEauthorblockA{\textit{Department of Electrical Engineering} \\
\textit{University of Twente}\\
Enschede, Netherlands \\
a.antonruiz@utwente.nl}
\and
\IEEEauthorblockN{Samar Hosseinzadegan,\\ John Kvarnstrand, Klas Arvidsson}
\IEEEauthorblockA{\textit{Bluetest AB}\\
Gothenburg, Sweden \\
name.familyname@bluetest.se}
\and
\IEEEauthorblockN{Andrés Alayón Glazunov}
\IEEEauthorblockA{\textit{Department of Science and Technology} \\
\textit{Linköping University}\\
Norrköping Campus, Sweden \\
andres.alayon.glazunov@liu.se}
}

\maketitle

\begin{abstract}
In this paper, we propose quantifying the radiated power of phased arrays or, in general, directive antennas, by the Constrained-View Radiated Power (CVRP). The constrained view shall be interpreted here as the Field-of-View (FoV) of an antenna that defines a region in space where focusing the radiated power is highly desired. In the limiting cases, we have that CVRP equals the Total Radiated Power (TRP) when the FoV covers the whole sphere, while, if the FoV reduces to a single point in space, the CVRP equals the Equivalent Isotropic Radiated Power (EIRP). We further present an analysis based on measured radiation patterns of a 16-element, linearly polarized, millimeter-Wave (mmWave), planar phased array antenna operating at 28 GHz. We compare the results to two ideal planar array antennas with the same number of Huygens and cosine elements. The evaluated figure of merit is computed for different scanning angles, as well as for different malfunctions of antenna elements, both for the real and simulated arrays. The results show that the introduced figure of merit could be potentially used for the detection of malfunctioning elements in antenna arrays as well as to characterize the impact of scan loss. Furthermore, CVRP is useful to straightforwardly and significantly characterize the performance of a directive antenna in terms of the power radiated towards a specific region in space. 
\end{abstract}

\IEEEpeerreviewmaketitle

\section{Introduction}
With the advent of \ac{5G}, the shift towards higher frequencies such as \ac{mmWave} has deemed phased arrays as a major enabling technology, helping to overcome propagation loss thanks to their large directivity and ability to direct the transmitted power towards the desired direction electronically, allowing real-time tracking. From the first prototypes to the final products, \ac{OTA} testing is paramount to assess the performance of antennas in general. \ac{OTA} testing, especially for phased array antennas and any other directive antennas, includes, among other antenna characteristics, the measurement of the radiation pattern of the \ac{AUT}, e.g., in terms of the angle-dependent \ac{EIRP} or just its maximum. From the angle-dependent \ac{EIRP}, another \ac{FoM} that can be extracted is the \ac{TRP}, which 
characterizes the radiated power integrated over the whole sphere. \ac{OTA} \ac{TRP} measurement challenges for \ac{5G} systems have been addressed in the literature, see, e.g., \cite{TRPOTA}. Modifications to the \ac{TRP} have arisen in the literature in order to characterize the radiation toward specific, more focused, angular regions. This is because, in many applications, directive antennas are employed. For example, the \acp{PRP} has been introduced \cite{5GAA,ILMVG}, and is defined as
\begin{equation}
\label{Eq1}
PRP=\frac{1}{4\pi}\int_{0}^{2\pi}\int_{\theta_1}^{\theta_2}{EIRP\left(\theta,\varphi\right)\sin{\left(\theta\right)}d\theta d\varphi},
\end{equation}
where $\theta_1$ and $\theta_2$ will be variable depending on the application. For example, for automotive \ac{GNSS}, $\theta_1=0^{\circ}$ and $\theta_2=90^{\circ}$, thus defining the \ac{UHRP}, while for \ac{V2X} communications, \ac{N75PRP}, where $\theta_1=60^{\circ}$ and $\theta_2=90^{\circ}$ or \ac{NHPRP}, where $\theta_1=60^{\circ}$ and $\theta_2=120^{\circ}$ are preferred.
From the \ac{PRP} definition, it follows that PRP~$\le$~TRP, being equal when $\theta_1=0^{\circ}$ and $\theta_2=180^{\circ}$.

While \ac{PRP} provides information regarding how much power is radiated towards a set of directions, it has a major drawback that, to the best of the authors' knowledge, has not been fully addressed in the literature. Indeed, the \ac{PRP} values computed for different spherical region sizes cannot be fairly compared since the normalization is performed for a constant area corresponding to the full sphere ($4\pi$). The implications of this can be observed with a simple example. Let us assume that we have an isotropic antenna, which, by definition, provides the same coverage in terms of \ac{EIRP} to any set of directions. If we were to use, for example, the \ac{UHRP} and \ac{N75PRP} \ac{FoM}, the result would be that \ac{UHRP}$>$\ac{N75PRP}, while, as previously stated, this isotropic antenna provides the same coverage in terms of \ac{EIRP} to both sets of directions. There is, however, an example where the angular area normalization is performed for the appropriate covered area \cite{Episphere}, although it is just for the specific case of having half sphere.

This paper introduces a new \ac{FoM}, the \ac{CVRP}, which overcomes this limitation and is suitable to assess the coverage of an antenna towards a specific set of directions that  weighs in the \ac{FoV} covered by those directions. Similarly to \ac{PRP}, \ac{CVRP} also does not need a full acquisition of the \ac{EIRP} radiation pattern of the \ac{AUT}, which can have a significant impact on measurement times, depending on the \ac{FoV} of interest. In addition, if only the area of maximum \ac{EIRP} values is required to be measured, e.g., towards the beamforming direction, then some relaxations on the measurement distances to acquire \ac{EIRP} are possible, as shown in \cite{ShortFD}. 

\section{Definition of CVRP}
The \ac{CVRP} is given by 
\begin{equation}
\resizebox{.9\hsize}{!}{$CVRP=\oint{EIRP\left(\theta,\varphi\right)P\left(\theta-\theta_c,\varphi-\varphi_c\right)\sin{\theta}d\theta d\varphi},$}
\label{Eq2}
\end{equation}
where we have introduced the spherical masking function $P\left(\theta-\theta_c,\varphi-\varphi_c\right)$ that defines, in the most general sense, an angular coverage area, centered at the observation angles $\left(\theta_c,\varphi_c\right)$. The function satisfies the normalization to unity integral over the sphere of unit radius
\begin{equation}
\oint{P\left(\theta-\theta_c,\varphi-\varphi_c\right)\sin{\theta}d\theta d\varphi}=1.
\label{Eq3}
\end{equation}
It is worthwhile to note that when the spherical masking function is given by
\begin{equation}
P\left(\theta-\theta_c,\varphi-\varphi_c\right)=\frac{1}{4\pi},
\label{Eq4}
\end{equation}
which describes a scenario where all directions are equally important, i.e., the \ac{FoV} comprises the whole sphere, and
\begin{equation}
CVRP=TRP,
\label{Eq5}
\end{equation}
as expected, while in the other limiting case for which the \ac{FoV} is limited to a single direction in space, the masking function is given by
\begin{equation}
P\left(\theta-\theta_c,\varphi-\varphi_c\right)=\frac{\delta\left(\theta-\theta_c,\varphi-\varphi_c\right)}{\sin{\theta_c}},
\label{Eq6}
\end{equation}
where $\delta\left(\theta-\theta_c,\varphi-\varphi_c\right)$ denotes the Dirac’s delta function centered at observation angles $\left(\theta_c,\varphi_c\right)$. Then, as expected too, in this case, we get
\begin{equation}
CVRP=EIRP\left(\theta_c,\varphi_c\right).
\label{Eq7}
\end{equation}
Hence, the definition of \ac{CVRP} is consistent with the \ac{TRP} and the \ac{EIRP} in the well-known limiting cases. For the rest of the in-between cases, we have that the masking function will normalize by the  covered area, making that, in the aforementioned example of the isotropic antenna, \ac{CVRP} will be the same value no matter the chosen \ac{FoV}, and therefore \ac{CVRP} values from different \acp{FoV} can be compared as a \ac{FoM} of angular coverage. Consequently, another way to interpret this \ac{FoM} is that if an \ac{AUT} has a given value of \ac{CVRP} at a given considered area or \ac{FoV}, such value would be achieved by another antenna with a constant \ac{EIRP}, equal to the \ac{CVRP} value, across all the considered area or \ac{FoV}.

In this paper, the proposed \ac{FoM} is used to evaluate the performance of an active phased array with several different configurations, including beam steering, as well as for two numerically simulated arrays with similar characteristics to the real phased array, for the sake of comparison.

\section{Measurement and simulations setup and procedure}
\subsection{Phased array antenna}

\begin{figure}
\centering
\includegraphics[width=0.8\columnwidth]{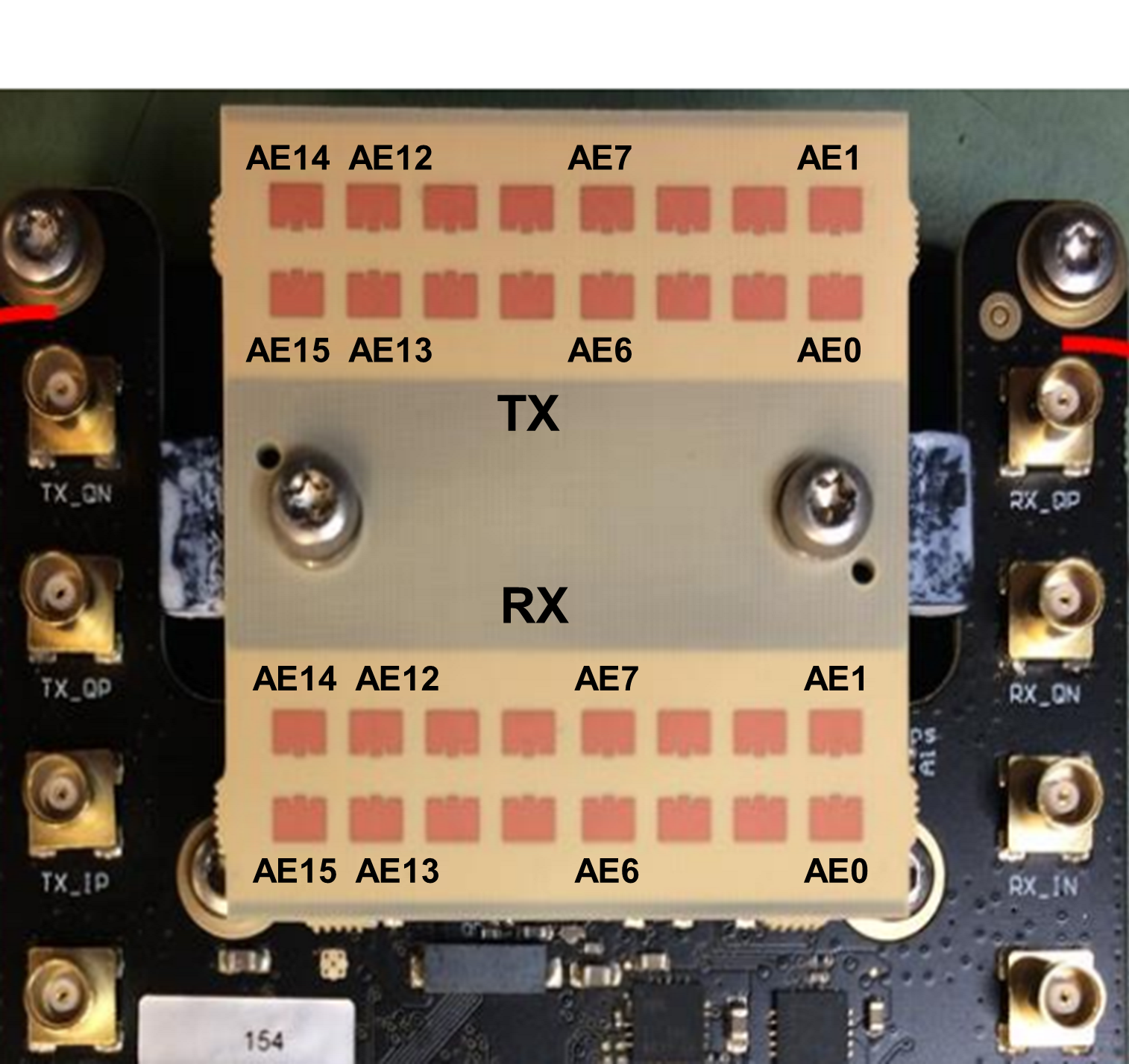}
\caption{Sivers Semiconductors AB EVK02001}
\label{F1}
\end{figure}

Fig.~\ref{F1} shows the active phased array measured and evaluated in this work. Depicted is an evaluation kit (EVK02001) with the \ac{RF} module (BFM02003) from Sivers Semiconductors AB \cite{Sivers}, which has its own continuous wave source. It operates from $24-29.5$~GHz. The \ac{RF} module is split into two identical modules (TX and RX), each with $2$ rows of $8$ linearly polarized patch \acp{AE}. Only the TX module was used for this study at the working frequency of $28$~GHz. The array antenna can perform scanning in the horizontal broadside plane, covering from $-45^{\circ}$ to $+45^{\circ}$ with $4.5^{\circ}$ step size. Therefore, it produces 21 different beams, which we denote as beam 1 starting from $-45^{\circ}$ and so on.

For this work, only beams 1 ($-45^{\circ}$), 10 ($-4.5^{\circ}$) and 11 ($0^{\circ}$) were considered.
In addition, to emulate the impact of faulty elements (FE), some were also switched off, both individually and in groups of two. It is worthwhile to note that in doing so, only the complete lack of feeding to the faulty elements has been evaluated. In particular, and taking as a reference the nomenclature of Fig.~\ref{F1}, elements (\ac{AE} in Fig.~\ref{F1}) 1, 3, 5, and 7 from the TX module were switched off individually, and these cases are referred as "FE 1", "FE 3", "FE 5" and "FE 7", respectively. Then, the pairs of elements 8\&7, 14\&1, 14\&7, and 15\&1 were switched off, being these referred to as "FE 8\&7", "FE 14\&1", "FE 14\&7", and "FE 15\&1", respectively. The cases where no elements were switched off are called "All ON."
Therefore, between the 3 different beams and the 9 different element activation configurations, 27 different experimental radiation patterns were measured for this work.

\subsection{Simulations}

For the sake of reference, two arrays were simulated using the MATLAB Phased Array System Toolbox. They have a very similar structure to the Sivers phased array. Thus, they have $2$ rows of $8$ elements each, with an \ac{IES} of $\lambda/2$ and a working frequency of $28$~GHz. Note that the \ac{IES} of the Sivers phased array at $28$~GHz is not exactly $\lambda/2$, although it is not far from it, so the results should be comparable. The elements are considered to be cosine \ac{AE} for one of the arrays and Huygens sources (cardioids) in the other case. Therefore, the cosine array will be referred to as "CS," and the one made up of Huygens sources as "HS." Beam steering has also been performed for both arrays in order to match the beam steering of the Sivers array. In addition, a simulation of the failure of elements 14\&7 of the array was carried out. The results from the MATLAB Phased Array System Toolbox are the antenna gains for each of the simulated arrays, which, since no losses were considered, are equivalent to their directivity. Therefore, to obtain the pattern in terms of \ac{EIRP}, the following equation was used
\begin{equation}
EIRP(\theta,\varphi)\mathrm{[dBm]}=TRP\mathrm{[dBm]}+D(\theta,\varphi)\mathrm{[dBi]},
\label{Eq8}
\end{equation}
where the \ac{TRP} is the one corresponding to the Sivers antenna with the same beam steering angle, and all elements are switched on, denoted as "All ON." The \ac{TRP} is computed as follows \cite{CTIA}
\begin{equation}
\resizebox{.99\hsize}{!}{$TRP\cong\frac{\Delta\varphi\Delta\theta\ }{4\pi}\sum_{i=1}^{N-1}\sum_{j=0}^{M-1}{\left[EIRP_\theta\left(\theta_i,\varphi_j\right)+EIRP_\varphi\left(\theta_i,\varphi_j\right)\right]\sin{\left(\theta_i\right)}},$}
\label{Eq9}
\end{equation}
where $N=120$ is the number of sampling points of $\theta$, $M=240$ is the number of sampling points of $\varphi$. $\Delta\varphi$ and $\Delta\theta$ are both equal to $0.0083\pi$ radians ($1.5^{\circ}$) since both simulations and experimental measurements were performed with a $1.5^{\circ}$ step size in both $\varphi$ and $\theta$. Note that \ac{EIRP} here is split into two components with orthogonal polarizations $EIRP_\theta\left(\theta,\varphi\right)$ and $EIRP_\varphi\left(\theta,\varphi\right)$, since that is how experimental measurements were acquired. The relationship between \ac{EIRP} and its orthogonal polarizations components is defined by
\begin{equation}
\resizebox{.89\hsize}{!}{$EIRP\left(\theta,\varphi\right)=EIRP_\theta\left(\theta,\varphi\right)+EIRP_\varphi\left(\theta,\varphi\right).$}
\end{equation}
Note that the summation is performed in linear units, e.g. mW, and also that, unless explicitly stated with the use of $\mathrm{[dBm]}$, all \ac{EIRP} and \ac{CVRP} found in equations are expressed in linear units.
\subsection{Measurements}

The \ac{OTA} measurements were performed with an RTS65 \ac{RC} from Bluetest AB. The RTS65 used for this work is equipped with the \ac{CATR} option \cite{CATR,BluetestCATR}, which enables radiation pattern measurements inside the \ac{RC}. Reflections are handled with a series of patented frequency selective absorbers, which behave as absorbers for FR2 frequencies, to which the working frequency of $28$~GHz used in this work belongs. A Gregorian dual reflector system with numerically shaped surfaces provides a cylindrical quiet zone of $30$~cm of diameter, enough to accommodate the Sivers antenna, with $0.6$~dB amplitude ripple STD, and $4^{\circ}$ phase ripple STD, at a frequency range from $24.25-42$~GHz. Two orthogonal polarizations are measured with this system, thus acquiring $EIRP_\theta\left(\theta,\varphi\right)$ and $EIRP_\varphi\left(\theta,\varphi\right)$. The dynamic range of this system is over 20 dB, which is lower than that from other commercial CATR solutions that use an anechoic chamber, typically achieving dynamic ranges between 50 and 80~dB \cite{MVGCATR}. However, amplitude and phase ripples are in line with those other solutions.

\begin{figure}
\centering
\includegraphics[width=0.8\columnwidth]{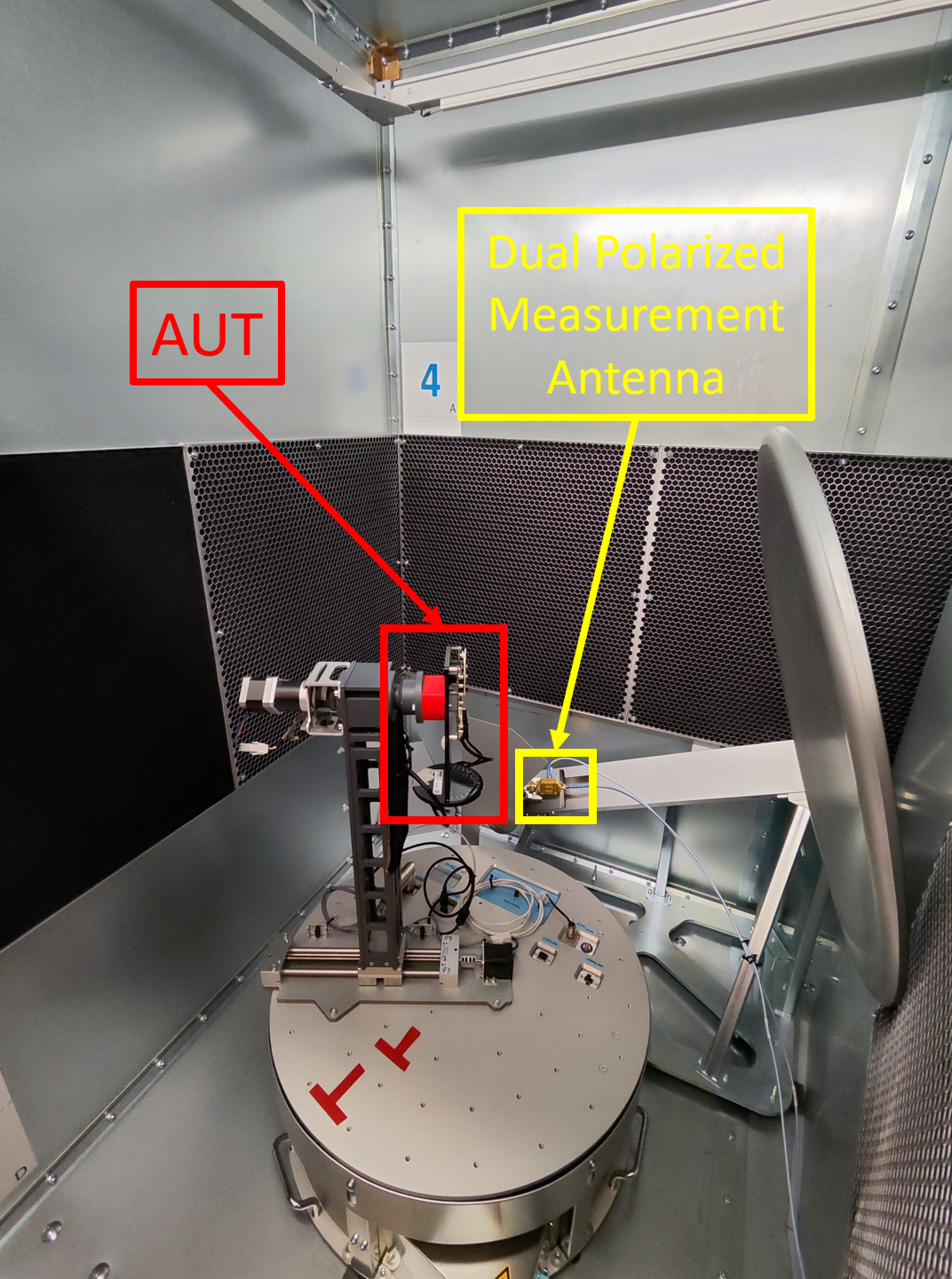}
\caption{RTS65 with CATR option and Sivers antenna}
\label{F2}
\end{figure}

\begin{figure}
\centering
\includegraphics[width=0.8\columnwidth]{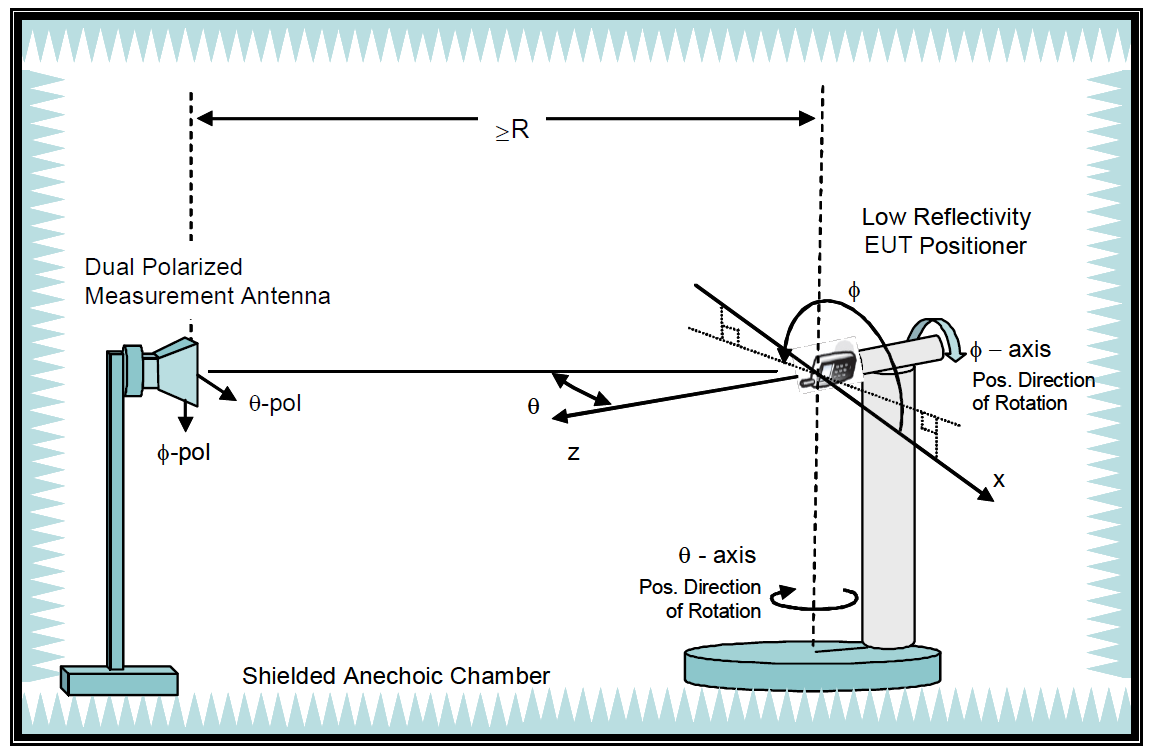}
\caption{Distributed-axes system. Source: \cite{CTIA}}
\label{F3}
\end{figure}

As shown in Fig.~\ref{F2}, the Sivers antenna is held in place by a roll tower over a turntable. This corresponds to a distributed-axes system, such as the one depicted in Fig.~\ref{F3}. The measured angles are $-171^{\circ}$ to $171^{\circ}$ in $\theta$ and $0^{\circ}$ to $180^{\circ}$ in $\varphi$, with $1.5^{\circ}$ step in both cases. For the angles that are not covered, which correspond to the back direction of the antenna, the \ac{EIRP} is assumed to be 0 (in linear units).

Since the antenna has its own continuous wave source, it is not possible to measure with a \ac{VNA}, so a spectrum and signal analyzer was used instead to measure the \ac{EIRP} of the antenna.

\subsection{Application of CVRP}

As expressed in (\ref{Eq2}), \ac{CVRP} is very similar to the equation for \ac{TRP}, being the difference in the multiplication of the masking function inside the integral. The masking function is governed by $\theta_c$, $\varphi_c$, $\theta_{FoV}$, and $\varphi_{FoV}$. Therefore
\begin{equation}
\theta_{min}=\theta_c-\frac{\theta_{FoV}}{2},
\label{Eq10}
\end{equation}
\begin{equation}
\theta_{max}=\theta_c+\frac{\theta_{FoV}}{2},
\label{Eq11}
\end{equation}
\begin{equation}
\varphi_{min}=\varphi_c-\frac{\varphi_{FoV}}{2},
\label{Eq12}
\end{equation}
\begin{equation}
\varphi_{max}=\varphi_c+\frac{\varphi_{FoV}}{2}.
\label{Eq13}
\end{equation}
Numerically, the implementation is based in (\ref{Eq9}). The application of the masking function is as follows: on the one side, all the values of $EIRP_\theta\left(\theta_i,\varphi_j\right)$ and $EIRP_\phi\left(\theta_i,\varphi_j\right)$ that are outside of the intervals $[\varphi_{min},\varphi_{max}]$ and $[\theta_{min},\theta_{max}]$ are set to 0, while the rest remain unchanged, obtaining $EIRP_{\theta,msk}\left(\theta_i,\varphi_j\right)$ and $EIRP_{\varphi,msk}\left(\theta_i,\varphi_j\right)$. On the other hand, (\ref{Eq9}) is modified by substituting $4\pi$, which corresponds to the area of the whole unit sphere, by the area $A$ that falls within the $[\varphi_{min},\varphi_{max}]$ and $[\theta_{min},\theta_{max}]$ intervals. This area is computed from the analytic solution of the equation 
\begin{equation}
A=\int_{\varphi_{min}}^{\varphi_{max}}\int_{\theta_{min}}^{\theta_{max}}{r^2\sin{\left(\theta\right)}d\theta d\varphi},
\label{Eq14}
\end{equation}
where $r=1$ since the considered sphere is the unit sphere. Therefore, we have that
\begin{equation}
\resizebox{.99\hsize}{!}{$CVRP\cong\frac{\Delta\varphi\Delta\theta\ }{A}\sum_{i=1}^{N-1}\sum_{j=0}^{M-1}{\left[EIRP_{\theta,msk}\left(\theta_i,\varphi_j\right)+EIRP_{\varphi,msk}\left(\theta_i,\varphi_j\right)\right]\sin{\left(\theta_i\right)}},$}
\label{Eq15}
\end{equation}
where the variables entering the summations are given above and, as for (\ref{Eq9}), $N=120$, $M=240$, and $\Delta\varphi$ and $\Delta\theta$ are both equal to $0.0083\pi$ radians ($1.5^{\circ}$). In the case when $\theta_{FoV}$ and $\varphi_{FoV}$ are 0 the computation simplifies to
\begin{equation}
CVRP=EIRP_\theta\left(\theta_c,\varphi_c\right)+EIRP_\varphi\left(\theta_c,\varphi_c\right),
\label{Eq16}
\end{equation}
i.e., \ac{CVRP} equals the \ac{EIRP} of the observation point.

\section{Results}
\subsection{Considered CVRP cases}

\begin{figure}
\centering
\includegraphics[width=0.8\columnwidth]{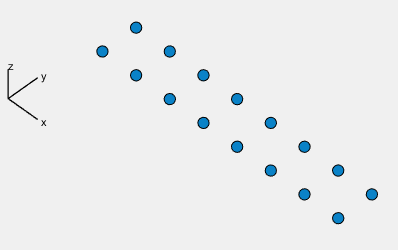}
\caption{Array orientation. Note that the front of the array looks towards the positive z-axis}
\label{F4}
\end{figure}

For the sake of comparability, it has been decided to use the same \ac{CVRP} sets of directions or \acp{FoV} for all the experimental and simulation data. Therefore, the selected \acp{FoV} correspond to those of different spherical caps (note that the antenna is oriented perpendicularly to the z-axis, as depicted in Fig.~\ref{F4}, and as it can be observed in Fig~\ref{F2} and Fig~\ref{F3}). Those spherical caps cover the whole $\varphi$ range. As for $\theta_{FoV}$, it goes from the whole sphere or $180^{\circ}$, then $165^{\circ}$, $150^{\circ}$, $135^{\circ}$, $120^{\circ}$, $105^{\circ}$, $90^{\circ}$, $60^{\circ}$, $45^{\circ}$, $30^{\circ}$, $21^{\circ}$, $15^{\circ}$, $9^{\circ}$, $6^{\circ}$, $3^{\circ}$, and, finally $\theta_{FoV}$ and $\phi_{FoV}$ both equal to $0^{\circ}$, which corresponds to the \ac{EIRP} value for $\theta$ and $\varphi$ both equal to $0^{\circ}$. Note that, in all cases, $\theta_c$ and $\varphi_c$ equal $0^{\circ}$.

The problem with this approach is that, while it provides a good insight into the coverage towards the desired direction for the broadside beam ($0^{\circ}$ scan angle), it does not have the same meaning for the other two considered beams. Therefore, the \ac{EIRP} patterns of beams 1 and 10 ($-45^{\circ}$ and $-4.5^{\circ}$ scan angles, respectively) are rotated around the y-axis so that the intended scan direction is aligned with the z-axis. This has the upside that makes comparing \ac{CVRP} values among different beams possible. However, the downside is that the y-axis rotation eliminates the equispacing of $\theta$ and $\varphi$ values. Therefore, it is needed to interpolate the \ac{EIRP} data to recover the $\theta$ and $\varphi$ equispacing, which is required to apply (\ref{Eq15}) \cite{CTIA}. The interpolation is linear and performed over the \ac{EIRP} data in linear units.

\subsection{CVRP comparison}

\begin{figure}
    \centering
        \subfloat[]{\includegraphics[width=1\columnwidth]{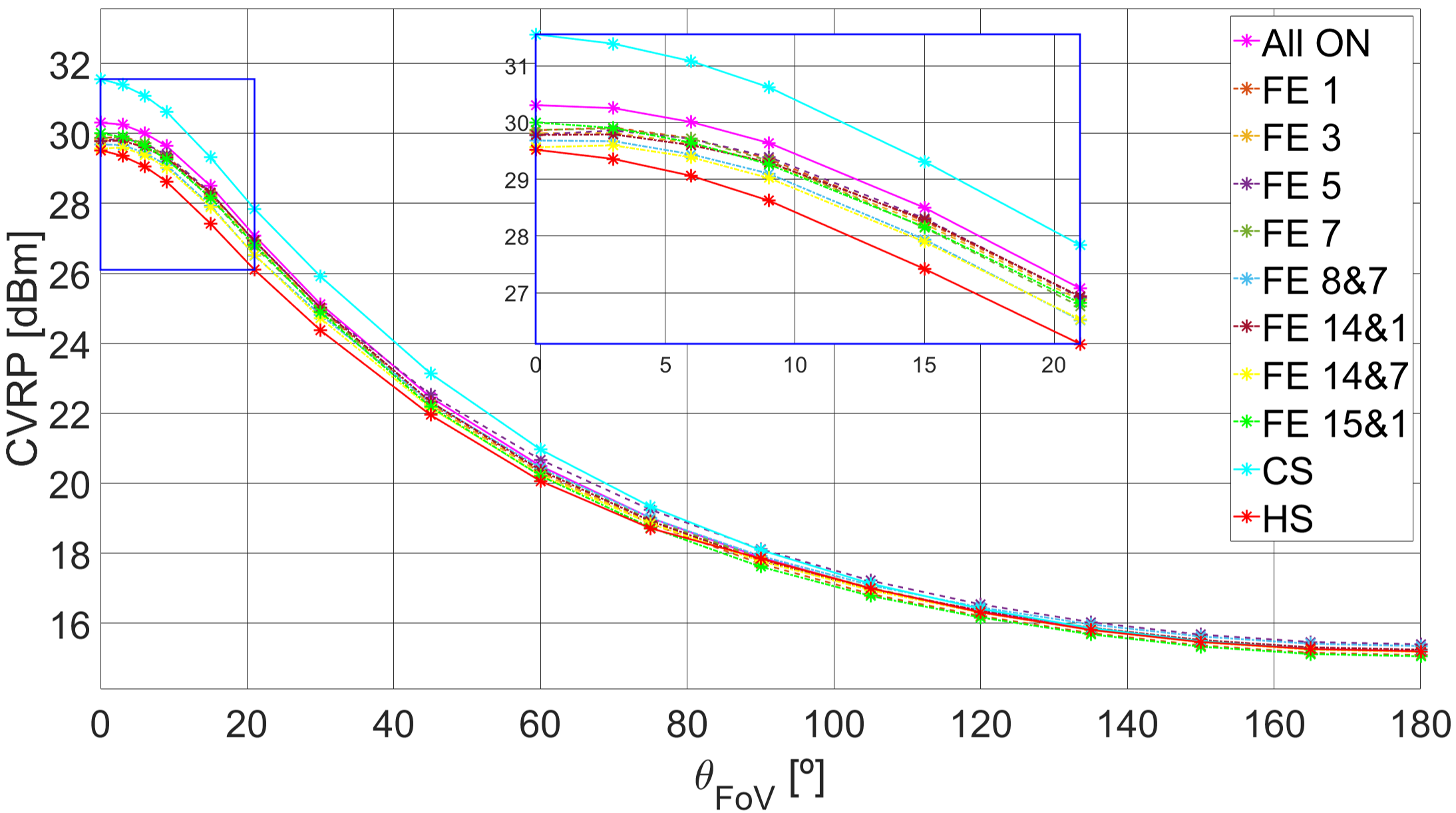}}
        \label{F5a}
        \subfloat[]{\includegraphics[width=1\columnwidth]{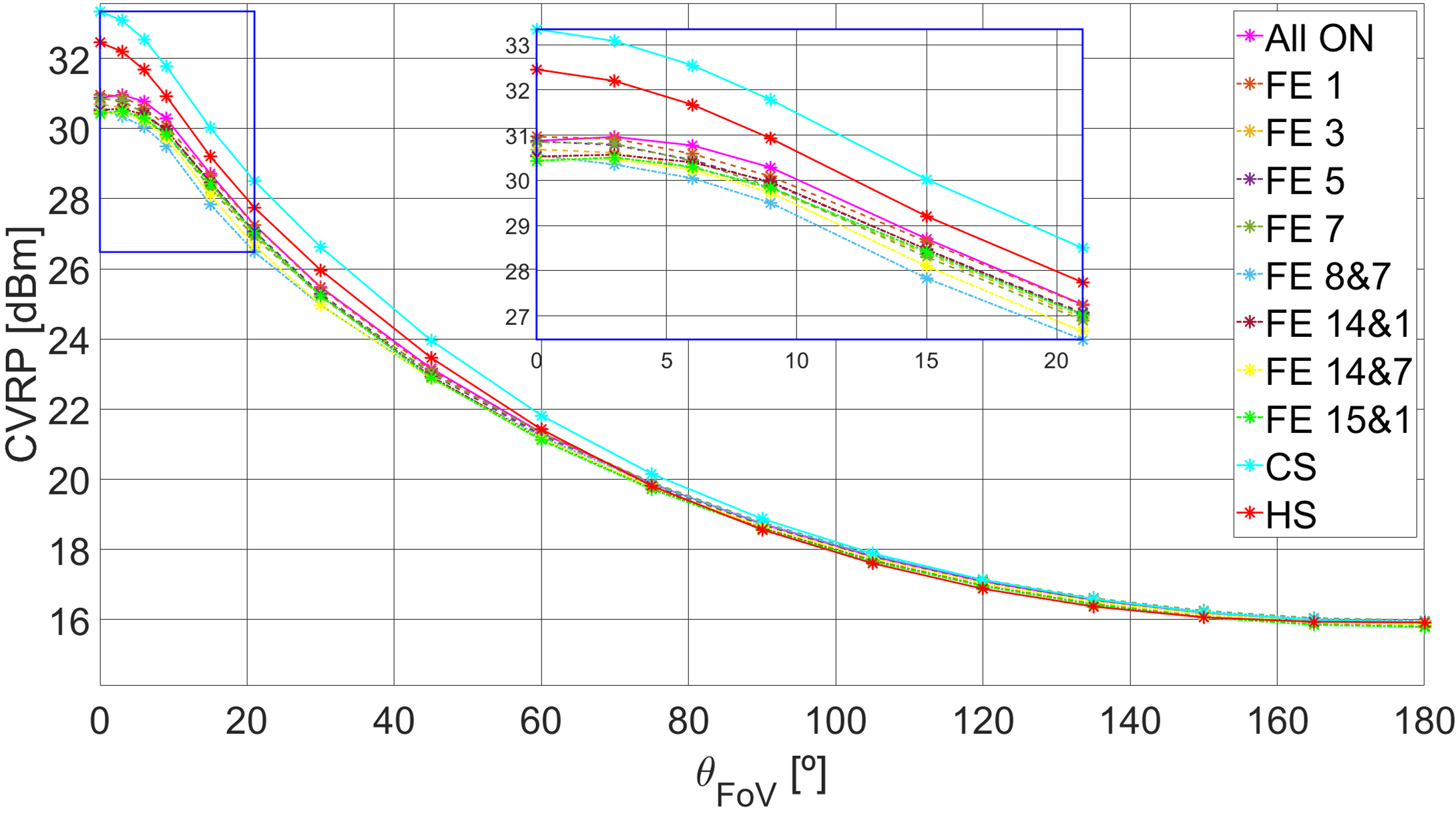}}
        \label{F5b}
        \subfloat[]{\includegraphics[width=1\columnwidth]{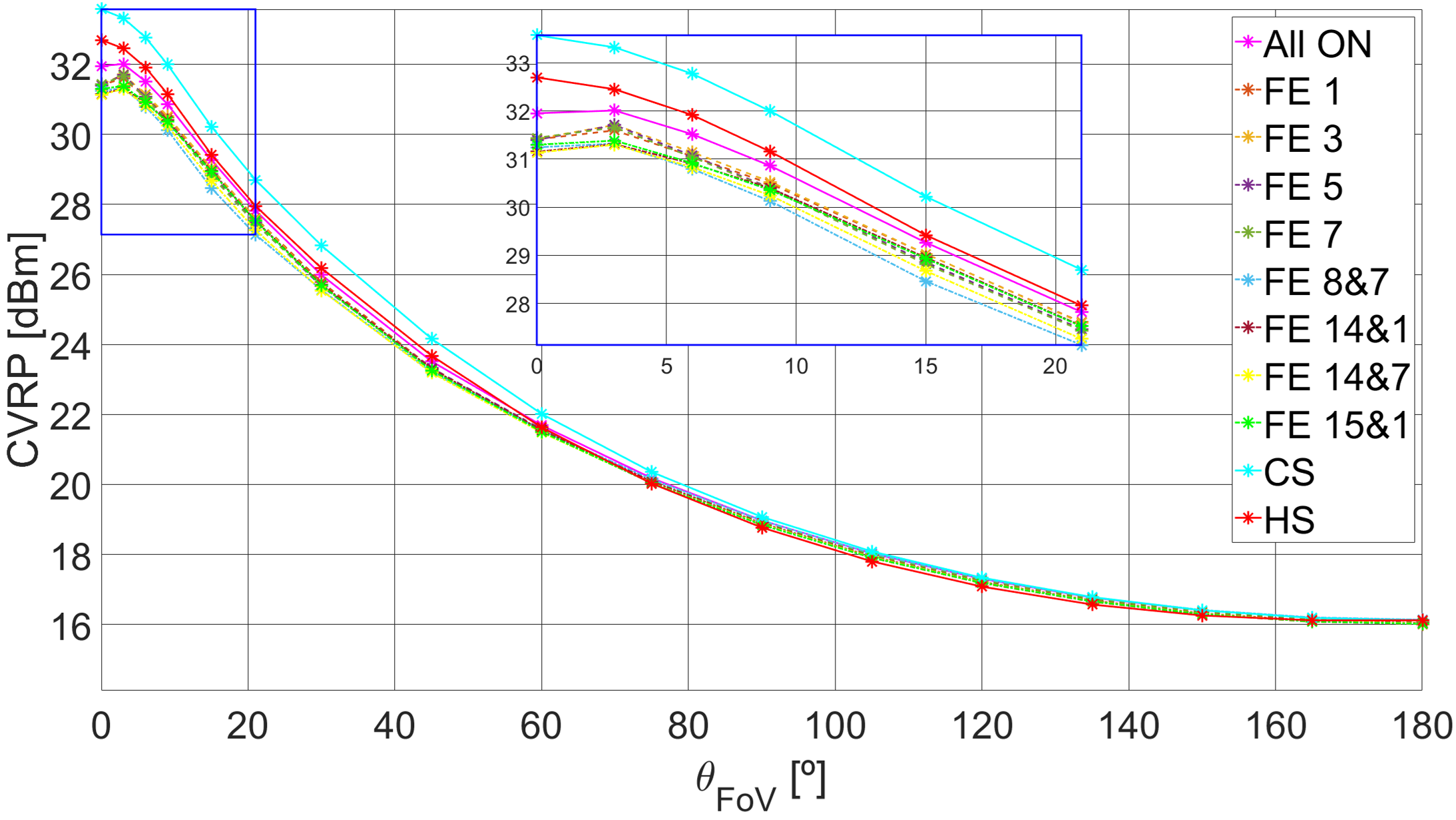}}
        \label{F5c}
    \caption{CVRP results as a function of $\theta_{FoV}$ (a) Beam 1, $-45^{\circ}$ scan angle, (b) Beam 10, $-4.5^{\circ}$ scan angle, (c) Beam 11, $0^{\circ}$ scan angle.}
    \label{F5}
\end{figure}

The \ac{CVRP} values for all the considered cases are shown in Fig.~\ref{F5}. Firstly, it can be observed that there is a general monotonic decrease of the \ac{CVRP} values from the \ac{EIRP} in the desired direction to the full-sphere \ac{CVRP} or \ac{TRP}. This implies that the antenna behaves as expected from a directive antenna, i.e., the antenna is focusing as much power as possible towards a narrow \ac{FoV} at the desired direction. Note that, from the definition of \ac{CVRP}, it follows that the \ac{CVRP} value will remain rather constant with larger \acp{FoV} whenever the \ac{EIRP} values in the new covered area are rather similar to the \ac{CVRP} of the area with the smaller \ac{FoV}. Note also that, in the extreme case of the ideal isotropic antenna, the \ac{CVRP} is constant for all \acp{FoV}, and also that, in case it was possible to have a beam that had constant \ac{EIRP} over a given \ac{FoV}, then what we would observe in the \ac{CVRP} values is that they would be constant up to that \ac{FoV}, and then they would decrease. In addition, in case of having increasing \ac{CVRP} values with increasing \acp{FoV}, it would imply that the antenna covers better in terms of \ac{EIRP} the wider area than the narrower one, which is not generally a desired result unless required for a specific application. In short, a monotonically decreasing \ac{CVRP} with increasing \acp{FoV} is expected from a well-pointed directive antenna, similarly to the results shown in Fig.~\ref{F5}.

On the other hand, the \ac{CVRP} values for the cosine simulated array ("CS") are generally larger than for the Huygens array ("HS") and the real array with all elements switched on ("All ON"), despite having the same \ac{TRP}. This result is expected when comparing the cosine array against the Huygens array due to the less directive radiation pattern of Huygens sources. As for the Huygens array, it has generally larger \ac{CVRP} values than the "All ON" case for scan angles of $0^{\circ}$ and $-4.5^{\circ}$, while being generally lower for the $-45^{\circ}$ scan angle. This might be due to the differences in the radiation patterns of both arrays, which might impact how much scan loss the beam steering introduces and how much distortion the rotation and interpolation introduce. In any case, it is clear that the performance of the real array ("All ON"), in terms of \ac{CVRP}, is closer to the Huygens array ("HS") than to the cosine array ("CS").

\begin{figure}
    \centering
        \subfloat[]{\includegraphics[width=1\columnwidth]{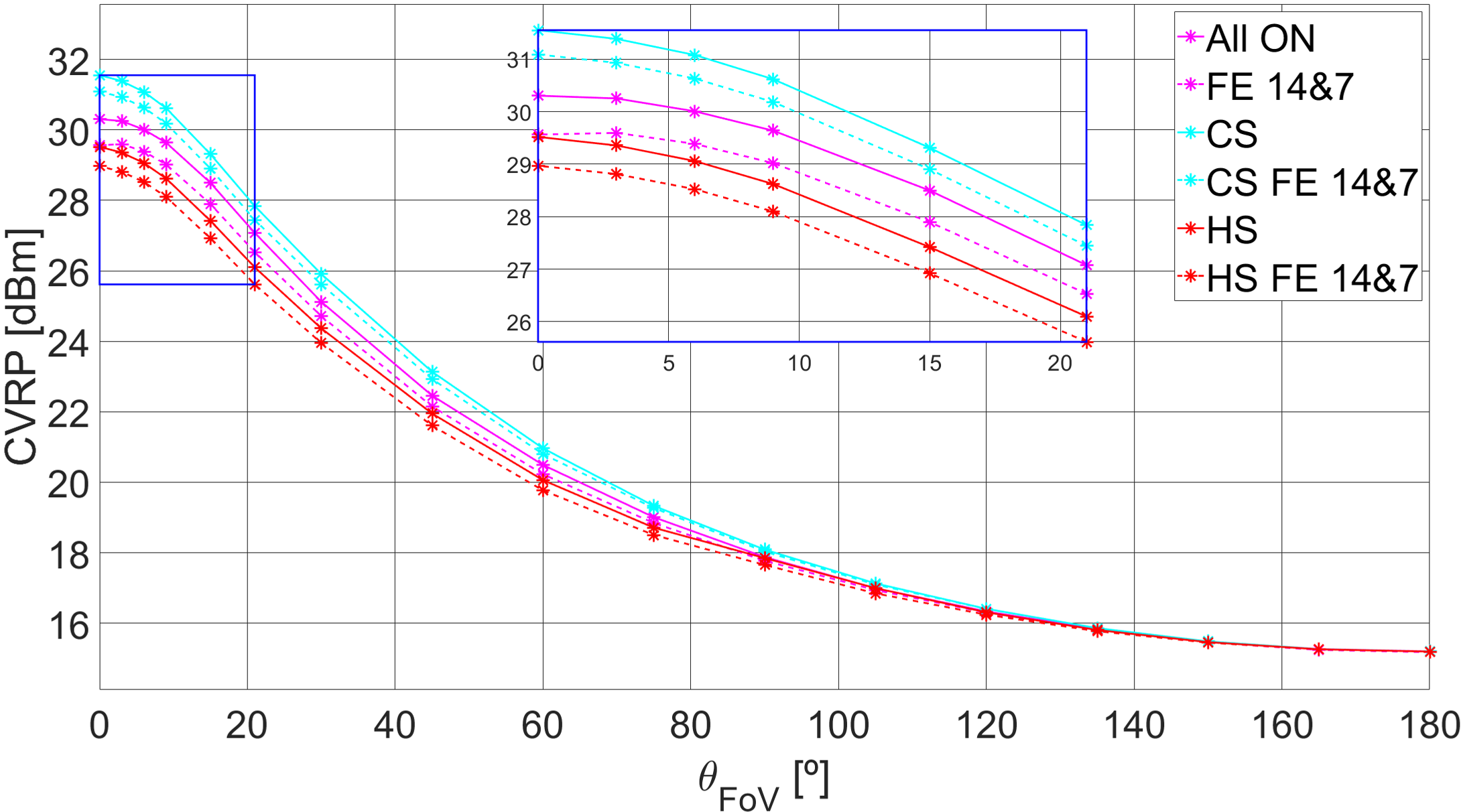}}
        \label{F6a}
        \subfloat[]{\includegraphics[width=1\columnwidth]{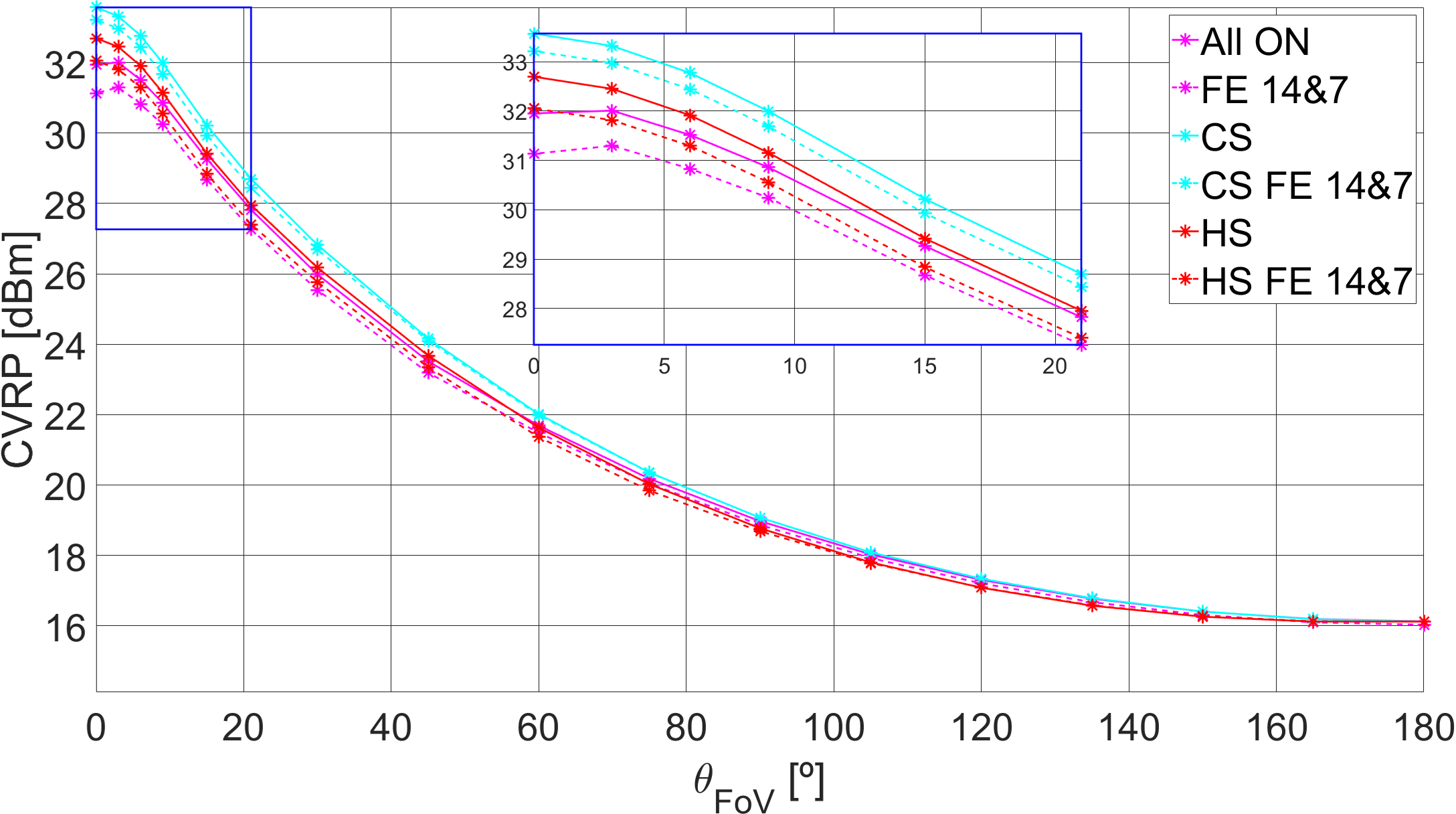}}
        \label{F6b}
    \caption{CVRP results as a function of $\theta_{FoV}$ for experimental and simulated \ac{AE} failures (a) Beam 1, $-45^{\circ}$ scan angle, (b) Beam 11, $0^{\circ}$ scan angle.}
    \label{F6}
\end{figure}

Moving now to one of the most interesting results, on the one hand, it can be observed the \ac{CVRP} values for large \acp{FoV}, including \ac{TRP}, are very similar for all the real array measurements where 1 or 2 elements have been switched off. On the other hand, differences are observed at narrow \acp{FoV}. This can be appreciated in the inset plots in Fig.~\ref{F5}. To confirm that these differences are expected when switching elements off, the "FE 14\&7" case was replicated in both simulated arrays, keeping the same \ac{TRP} output since it is what happens with the real array measurements. The results for beams 1 and 11 are shown in Fig.~\ref{F6}, where it can be seen that the differences in \ac{CVRP} values are larger for the narrower \acp{FoV} for all cases, including the real array ("All ON" and "FE 14\&7"), the Huygens array ("HS" and "HS FE 14\&7"), and the cosine array ("CS" and "CS FE 14\&7"). These differences are due to the more irregular beam shape when elements are switched off, which becomes a worse coverage of the desired area in terms of \ac{EIRP} or \ac{CVRP} values. Therefore, \ac{CVRP} might be useful for detecting malfunctioning \acp{AE} in arrays by measuring the radiated power withing a limited coverage area.

The fact that \ac{TRP} values for all real array measurements are very similar might be because switching one or two elements does not impact the total power output of the array, i.e., that when one element is switched off, other elements radiate more power and the total power output remains the same. Another possibility is that this is caused by the measurement uncertainty for \ac{TRP} of the CATR system or the dynamic range of the system, although this would be needed to be carefully addressed and tested for this particular case. However, neither of these hypotheses has been confirmed and is beyond the scope of this work.

Finally, it can be observed how the scan loss affects not only the maximum \ac{EIRP} (or \ac{CVRP} for $0^{\circ}$ \ac{FoV}) but also the \ac{CVRP} values, increasingly as the \ac{FoV} narrows down. This can be clearly seen when comparing Fig.~\ref{F5} (a) and (c), where a shift in the curve of \ac{CVRP} for steering angles of $-45^{\circ}$ and $0^{\circ}$ occurs, starting as a large shift for the narrow \acp{FoV} and converging to the same values for wide \acp{FoV}. This shows the effect of scan loss in the performance in terms of \ac{EIRP} coverage or \ac{CVRP}, which is more relevant for narrow \acp{FoV}. In any case, the comparison of both steering angles is subject to the distortion introduced when rotating and interpolating the radiation pattern of the $-45^{\circ}$ steering angle case.

As a final remark, it is worthwhile noting that the \ac{CVRP} can be measured with other \ac{OTA} systems that can generate a plane wave, e.g., in \cite{MSLildal2021} and \cite{Iupikov2022}. Moreover, the spherical masking can be done directly with the method provided in \cite{Iupikov2022}. Results based on these investigations will be the subject of future work.

\section{Conclusions}

This paper has introduced a novel \ac{FoM} to characterize directive antennas and how they radiate towards the desired directions in an alternative manner to the \ac{PRP} approach, allowing the comparison across different \acp{FoV}. The proposed \ac{FoM} is consistent with both the \ac{EIRP} and the \ac{TRP}. It is worth noting that a constant \ac{CVRP} is desirable within the \ac{FoV} of the antenna. Moreover, for highly directive antennas with narrow beams, the \ac{CVRP} will decrease as the considered \ac{FoV} increases beyond the beamwidth of the antenna, as demonstrated by our results.

Furthermore, the presented results have shown that this \ac{FoM} can be used to diagnose malfunctioning \acp{AE} in arrays producing narrow beams for narrow \acp{FoV}, as confirmed by the similar behavior of real measurements and simulations. In addition, this work shows the impact of scan loss of different beams in terms of the proposed \ac{FoM}, which will be higher for narrower \acp{FoV}. 

Further studies shall consider the error introduced by the rotation around the y-axis and the posterior interpolation, which is necessary for the fair comparison of different beams. In addition, the differences between the experimental results for arrays with all antenna elements on and with one or two faulty elements are small for large values of the \ac{FoV}. Since experimental data comes from an active \ac{AUT}, a careful assessment of the uncertainty budgets is necessary to determine whether the observed differences for the considered cases are statistically significant in order to fully establish the practical validity of the method to detect malfunctioning \acp{AE} in arrays.

The results shown by both experimental and simulated data showcase the potential of using this \ac{FoM} to characterize the performance of directive antennas with narrow beams from an angular coverage area standpoint.

\section*{ACKNOWLEDGEMENT}
The work of Alejandro Antón was conducted within the ITN-5VC project, which is supported by the European Union’s Horizon 2020 research and innovation program under the Marie Skłodowska-Curie grant agreement No. 955629. Andrés Alayón Glazunov also kindly acknowledges funding from the ELLIIT strategic research environment (https://elliit.se/).

\bibliographystyle{IEEEtran}

\bibliography{References}

\end{document}